\documentclass[fleqn,usenatbib]{mnras}

\usepackage{newtxtext,newtxmath}

\usepackage[T1]{fontenc}
\usepackage{ae,aecompl}
\usepackage{longtable}

\usepackage{graphicx}	
\usepackage{amsmath}	
\usepackage[flushleft]{threeparttable}

\usepackage{xcolor}

\usepackage{lscape}
\usepackage{ulem}

\title[A vanishing triple transient]{A bright triple transient that vanished within 50 minutes}
\author[E. Solano et al.]{
Enrique Solano$^{1}$\thanks{E-mail: esm@cab.inta-csic.es},
Geoffrey W. Marcy$^{2}$,
Beatriz Villarroel$^{3}$,
Stefan Geier$^{4,5}$,
\newauthor
Alina Streblyanska$^{5}$,
Gianluca Lombardi$^{4,5}$,
Rudolf E. Bär$^{6}$,
Vitaly N. Andruk$^{7}$,
\\
$^{1}$Centro de Astrobiolog\'{\i}a (CAB), CSIC-INTA, Camino Bajo del Castillo s/n, E-28692, Villanueva de la Ca\~{n}ada, Madrid, Spain\\
$^{2}$Space Laser Awareness, 3388 Petaluma Hill Rd, Santa Rosa, CA, USA 95404\\
$^{3}$Nordita, KTH Royal Institute of Technology and Stockholm University, Roslagstullsbacken 23, SE-106 91 Stockholm, Sweden\\
$^{4}$ Gran Telescopio Canarias (GRANTECAN), c/ Cuesta de San José s/n, E-38712 Breña Baja, La Palma, Spain\\
$^{5}$ Instituto de Astrofísica de Canarias, Avda Vía Láctea S/N, La Laguna, E-38205, Tenerife, Spain\\
$^{6}$ Institute for Particle Physics and Astrophysics, ETH Zurich, Wolfgang-Pauli-Strasse 27, CH-8093 Zurich, Switzerland\\
$^{7}$ Main Astronomical Observatory of the NAS of Ukraine, 27, Akademika Zabolotnoho St., Kyiv, 03143, Ukraine\\
}

\date{Accepted XXX. Received YYY; in original form ZZZ}

\pubyear{2022}

\begin{document}
\label{firstpage}
\pagerange{\pageref{firstpage}--\pageref{lastpage}}
\maketitle

\begin{abstract}
We report on three optically bright, $\sim$15th\,\rm{mag}, point-sources within 10\,\rm{arcsec} of each other that vanished within 1 hour, based on two consecutive exposures at Palomar Observatory on 1952 July 19 (POSS I Red and Blue).  The three point-sources have continued to be absent in telescope exposures during 71 years with detection thresholds of $\sim$21st\,\rm{mag}. We obtained two deep exposures with the 10.4-m Gran Telescopio Canarias on 25 and 27 April 2023 in $r$ and $g$-band, both reaching magnitude 25.5 (3-sigma). The three point-sources are still absent, implying they have dimmed by more than 10 magnitudes within an hour.  When bright in 1952, the most isolated transient source has a profile nearly the same as comparison stars, implying the sources are sub-arcsec in angular size and they exhibit no elongation due to movement. This triple transient has observed properties similar to other cases where groups of transients (``multiple transients'') have appeared and vanished in a small region within a plate exposure, \citep[e.g.,][]{Villarroel21, Villarroel22arxiv, Solano22}. The explanation for these three transients and the previously reported cases remains unclear. Models involving background objects that are optically luminous for less than one hour coupled with foreground gravitational lensing seem plausible. If so, a significant population of massive objects with structure serving as the lenses, to produce three images, are required to explain the sub-hour transients.
\end{abstract}

\begin{keywords}
astronomical data bases: surveys -- astronomical data bases: virtual observatory tools -- gravitational lensing: micro -- techniques: image processing.
\end{keywords}



\section{Introduction}
Since decades, astronomers have carefully scrutinized the sky looking for sources that vary in brightness. Numerous surveys have been conducted at all wavelengths, successfully revealing many classes of objects that vary by more than 1\,per cent. Low mass flare stars, pre Main Sequence objects, pulsating variables, novae, supernovae, gamma-ray bursts, fast radio bursts, tidal disruption events, and mergers of black holes and neutron stars are examples of some of these types of variable objects. 

However, only a few surveys have been carried out to detect fast transients (FT), defined as those that remain bright for less than one day. To name a few, \cite{Becker04}, using Deep Lens Survey data, reported three unusual optical transient events flaring on 1000\,\rm{s} timescales; \cite{Rau08} discovered 5 FTs (2 flaring M dwarfs and 3 periodic variables) using the Irénée du Pont telescope at Las Campanas, while \cite{Berger2013} found 19 FTs (8 asteroids and 11 flaring M dwarfs) using repeated observations of the Pan-STARRS1
Medium-Deep Survey (PS1/MDS) fields. Optical searches for transients lasting less than a minute were carried out by \cite{Richmond20} and \cite{Andreoni20}, but no such transients were definitively found.


Current searches for sub-hour transients suffer from confusion by tens of thousands of satellites and space debris in orbit around the Earth that reflect sunlight briefly and glint on time scales of seconds or less \citep[e.g.,][] {Corbett20, Nir21}. One solution to this problem would be to employ all-sky imaging performed prior to the launch of Sputnik in 1957 \citep{Villarroel22b}. The first Palomar Sky Survey (POSS I), with its images of the sky obtained in the early 1950’s, provides this database of images. The survey was conducted using the 48-inch
Oschin Schmidt telescope at Mount Palomar \citep{Minkowski63}, covers the entire sky north
of -45\textdegree $ $ declination and was carried out using photographic plates,
later converted into a digital format. In order to obtain colour information, each region of the sky was photographed twice, once using
a blue sensitive Kodak 103a-O plate, and once with a red sensitive
Kodak 103a-E plate, peaking at $\sim$4\,100 and $\sim$ 6\,400\,\rm{Å}, respectively.
The limiting photographic magnitudes of the blue and red plates are
21.1 and 20.0 mag, respectively. Typically, two exposures, one in the red and one in the blue, were taken within 1 hour of the same 6.5x6.5\textdegree $ $ field. 



The VASCO\footnote{\url{https://vascoproject.org/}}  team has  identified thousands of star-like sources that appeared on one photographic plate but not on the next one taken within 1 hour, constituting sub-hour transients\footnote{\url{http://svocats.cab.inta-csic.es/vanish/}}. The several thousand VASCO transient candidates have peak brightnesses of typically R magnitude 17 to 20, but they do not appear in later all-sky surveys with CCD detectors, indicating dimming by 2 to 5 mag.

Of particular interest was one field of 10\,arcmin that contained nine point-sources that did not appear in the image taken an hour later, nor in any images during the following 70 years \citep{Villarroel21, Villarroel22arxiv}. These “nine simultaneous” transients remain a mystery. Subsequent work identified 83 cases of such “multiple transients” including some that are apparently aligned as well as some double and triple transients \citep{Villarroel22arxiv}. 

In this \textbf{new} paper we describe an extraordinary transient candidate composed of three 15th mag point-like sources, within 10 arcsec of each other, that dimmed within an hour and for which deep imaging with the 10.4-m GTC telescope reveals no counterparts at a limiting magitude of 25th mag. 



\section{Observation of the triple transient}\label{observations}

We used the list\footnote{\url{http://svocats.cab.inta-csic.es/vanish-possi/}} of 5399 candidate point-source transients from \cite{Solano22} that were detected in digitised photographic exposures of the first Palomar Observatory Sky Survey (POSS I), and that are absent in images from modern surveys including Pan-STARRS, ZTF, and SDSS. In brief, SExtractor was used to identify sources in POSS I Red, and counterparts were searched for within a 5 arcsec radius in Gaia EDR3 and Pan-STARRS DR2 catalogues. Sources having no counterparts were retained for further vetting in other surveys, including POSS I B lue, POSS II, and ZTF. Those with no visible counterpart were deemed candidate transient sources.  

\begin{figure}
 \includegraphics[width=\columnwidth]{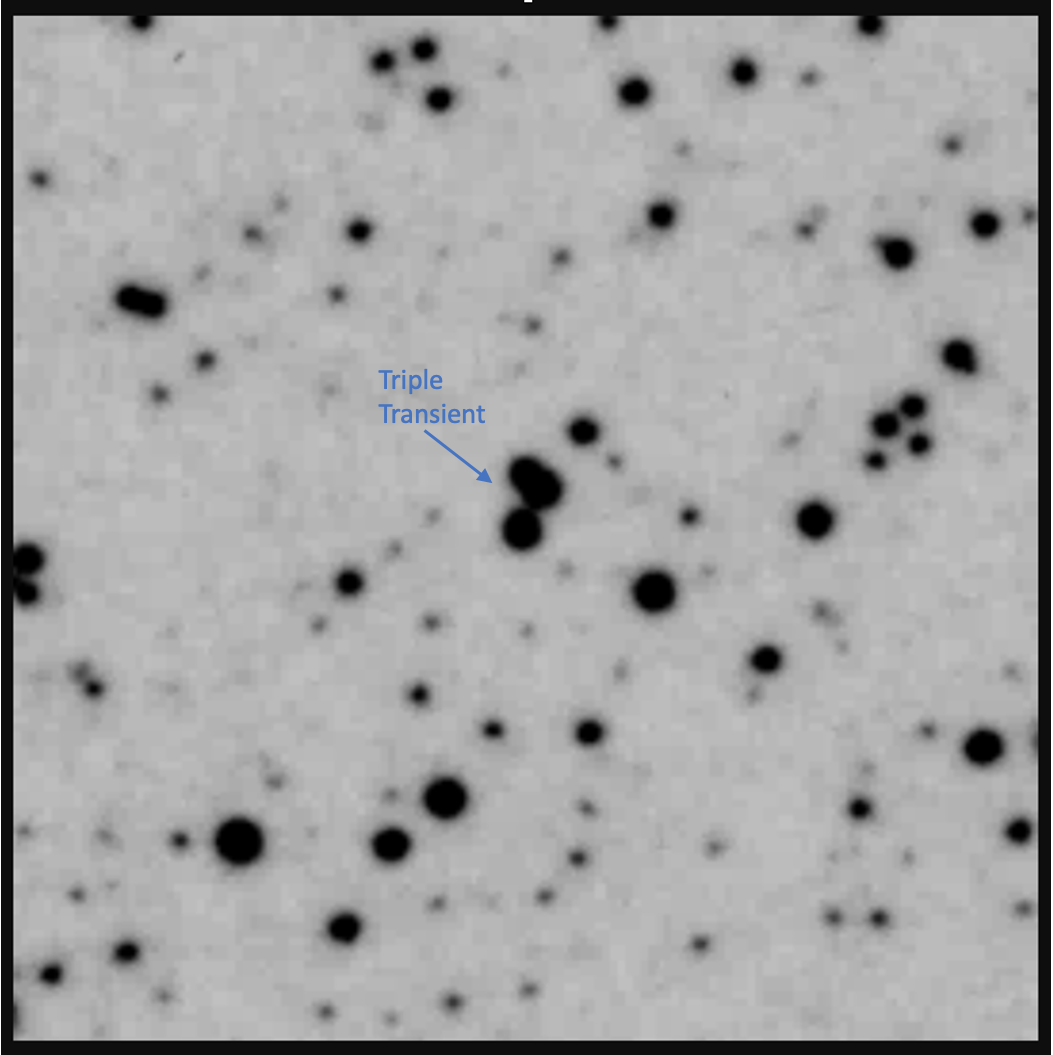}
    \caption{Supercosmos digitization \citep{Hambly01} of the Palomar Observatory Sky Survey (POSS) red-sensitive image, taken on 1952 July 19.  This is a zoom, 3x3\,\rm{arcmin}, centered on three bright ($R$ Supercosmos $\sim$ 15\,\rm{mag}) point-sources, marked by the blue arrow, at coordinates RA\,=\, 21$^{h}$18$^{m}$10.4$^{s}$, DEC\,=\,+50\textdegree22'43.4'' (J2000).  The three sources are absent in all subsequent exposures of this region during 71 years, including the POSS blue-sensitive image taken immediately after this exposure. The sources have profile shapes consistent with neighboring stars of similar brightness (Sect.\,\ref{analysis}).  They exhibit no evidence of peculiar shapes or elongation that would occur with asteroids, meteorites, cosmic rays, photographic plate defects, or aircraft.}
    \label{fig:SC}
\end{figure}

\begin{figure*}
\includegraphics[width=\textwidth]{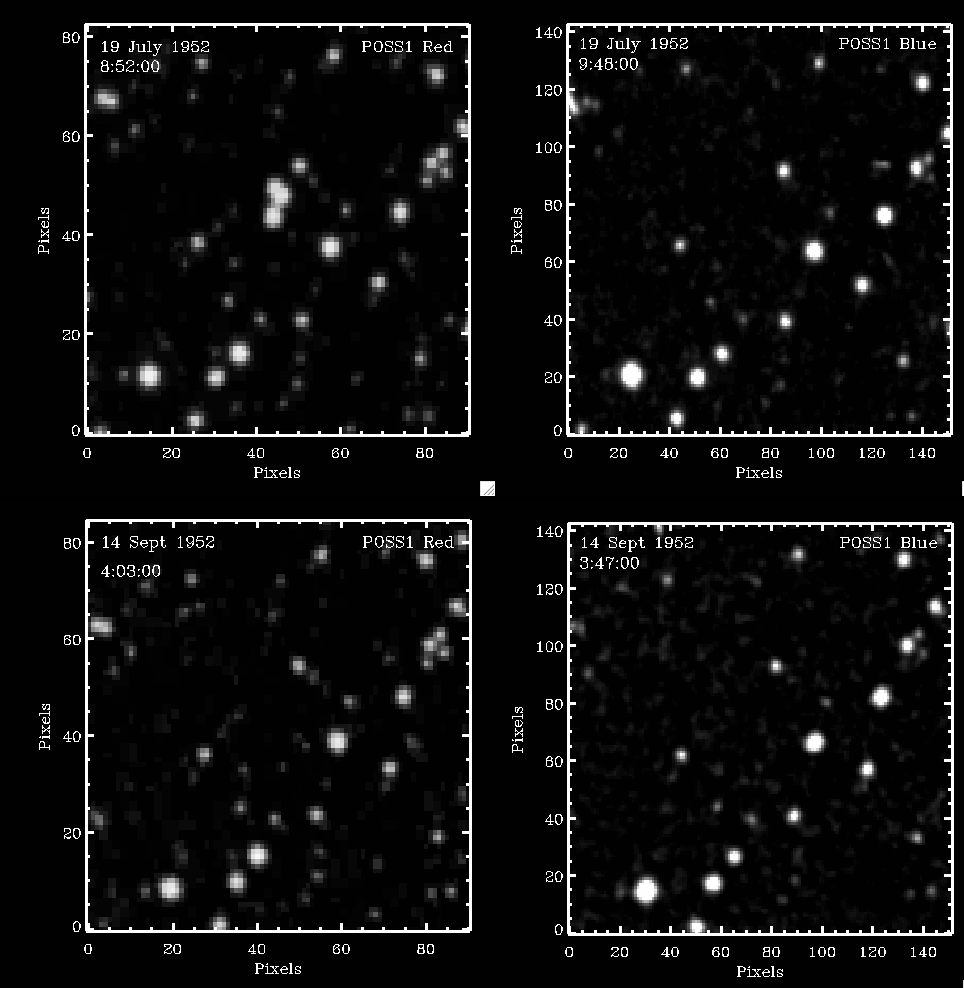}
\caption{Four exposures taken during 1952 of the 3x3\,arcmin region of sky centered on the triple transient seen in July 1952. Upper Left: the POSS I Red image on 1952 July 19 at 8:52 (UT) containing the triple transient just above center. Upper Right: a 10\,\rm{m} exposure POSS I Blue image taken immediately afterward, showing no evidence of the triple transient.  Lower Left: A POSS I Red image taken two months later (14 Sept.), showing the transient still gone.  Lower right: A POSS I Blue image on the same day showing the transient gone.  The triple transient dimmed by over 6 mag of its peak brightness within 50\,\rm{min} (or even more if the duration of the transient was shorter than the exposure time)  and remained undetected that year.}
\label{fig:stsci}
\end{figure*}

A particularly interesting case of multiple transient-event was discovered in the POSS I Red image, of 50\,\rm{min} exposure time, taken on 1952 July 19 (Fig.~\ref{fig:SC}). The center of the image contains the three star-like sources, identified by the blue arrow, at coordinates RA\,=\, 21$^{h}$18$^{m}$10.4$^{s}$, DEC\,=\,+50\textdegree22'43.4'' (J2000). Fig.~\ref{fig:SC}  shows the the POSS I photographic image digitised by SuperCosmos \citep{Hambly01} that provides high resolution and a sampling of 0.7\,arcsec per pixel instead of the usual 1.7\,arcsec per pixel in the STScI digitization of the POSS Digital Sky Survey. The three transients also appear clearly in the STScI digitization shown in Fig.~\ref{fig:stsci} at upper left.   

The three sources are roughly 15th mag in the $R$ Supercosmos band. The northern two sources are blended.  Eyeball inspection of the three transients in Fig.~\ref{fig:SC} reveals the individual sources to have profile shapes qualitatively similar to the neighboring stars. The point-like profiles are measured quantitatively in Sect.\,\ref{analysis}, showing they indeed exhibit no elongation or distortion compared to profiles of stars. Their star-like, circularly symmetric profile shapes distinguish them from the elongated and non-circular shapes of most asteroids, meteorites, aircraft, cosmic rays, photographic plate flaws, and any other moving objects during a 50\,\rm{m} exposure.

Fig.~\ref{fig:stsci} shows at upper left the STScI digitization of the three transients from the same POSS I red exposure.  The transients are less well sampled in this digitization, but they appear star-like, as in the SuperCosmos digitization. Immediately after the 50\,\rm{m} exposure, a 10\,\rm{m} POSS I blue exposure was taken, shown in the upper right of Fig.~\ref{fig:stsci}.  The three transient sources are absent. Two months later, in September, another red and blue pair of Palomar photographic exposures were taken of the same field, shown in lower left and lower right of the figure. The three transient sources are also absent.

\begin{figure*}
\includegraphics[width=\textwidth]{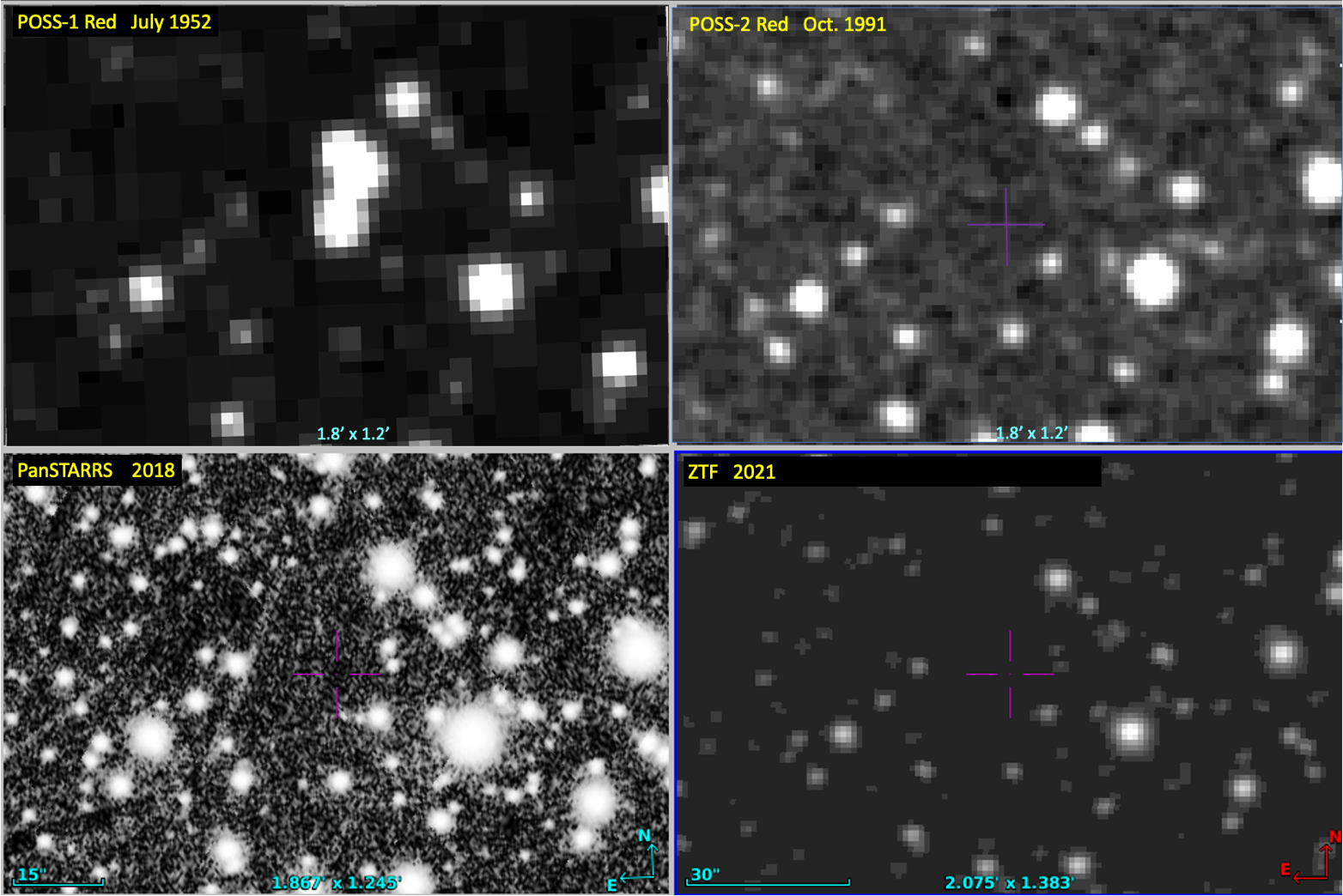}
\caption{Four exposures of the sky centered on the triple transient, taken at four different times during 69 years.  The field dimensions are approximately 1.8x1.2\,\rm{arcmin} and the position of the southernmost transient is marked by a purple cross.  Upper Left:  The POSS I red exposure, taken in July 1952, showing the triple transient composed of three $R$ Supercomos $\sim$ 15th mag point-lie sources.  Upper Right: The POSS II red exposure, taken in October 1991.  The triple transient is absent, with a detection threshold of 21st mag. Lower left:  The PanSTARRS exposure of the same field in 2018, showing no evidence of the triple transient.  Lower right: The ZTF exposure of the same field in 2021, showing the triple transient still undetected.  The upper two panels are STScI digitizations.}
\label{fig:comp}
\end{figure*}

We took advantage of the Virtual Observatory\footnote{http://ivoa.net} capabilities to look for the triple transient in more recent images and catalogues. The result of this search concluded that the transient does not appear in any later image of that region during the subsequent 69 years. Fig.\,\ref{fig:comp}, upper left, is the POSS I Red from 1952  showing the triple transient for reference.  The other three panels of the figure show images of the same field from POSS II red band taken in 1991, from Pan-STARRS in 2018, and from ZTF taken in 2021, all with limiting magnitudes roughly ~21.5.  None of those images exhibit the triple transient. A purple cross marks the location of the southernmost transient. 

\section{Observation of the triple transient with the 10.4m GTC telescope}
\label{GTC}
We used the OSIRIS imaging camera on the 10.4-m Gran Telescopio Canarias (GTC) to image the triple transient in both the $r$ and $g$-band on 25 and 27 April (UT) 2023 with total integration times of 30 min and 16.7 min, respectively. Both observations were obtained in dark time (without moon), with good seeing (<1"). The data reduction was done with self-written IDL codes. Standard bias-subtraction and flat-fielding were applied. The reduced images were aligned with integer pixel shifts (to avoid correlated noise in the combined images) and combined with a simple mean combination. The WCS information in the FITS headers was improved with the NOVA\footnote{\url{https://nova.astrometry.net/}} astrometry tool. We also used the Terapix swarp\footnote{\url{https://www.astromatic.net/software/swarp/}} procedure \citep{Bertin02} to improve the astrometry, using USNO-B1.0 astrometry as a reference field for zero-point calculations in the astrometrical solutions. We found that both methods give the same results within the expected errors. In this way, both GTC and POSS I images were calibrated in order to have the same astrometrical solution for both datasets. This approach allows us to directly compare source positions using the same reference catalogue. Despite the difference of 71 years between GTC and POSS images, we noticed that only a few objects show measurable proper motion while most of the stars in the field seem to have remained at the same position.

The resulting images have detection thresholds near 25.5 mag (3-sigma) at both $r$-and $g$-bands, 10 magnitudes fainter than the three transients were in the discovery image (Fig.\,\ref{fig:SC}).   

\begin{figure*}
\includegraphics[width=8.5cm]{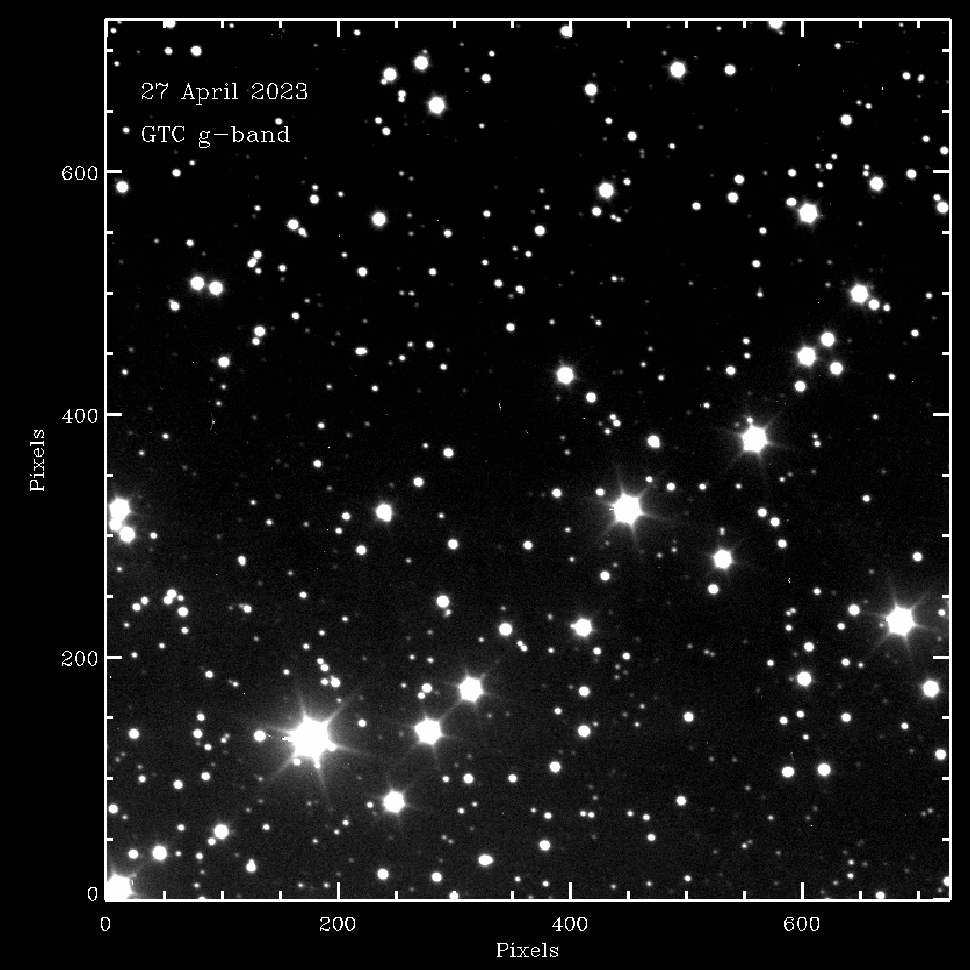}
\includegraphics[width=8.5cm]{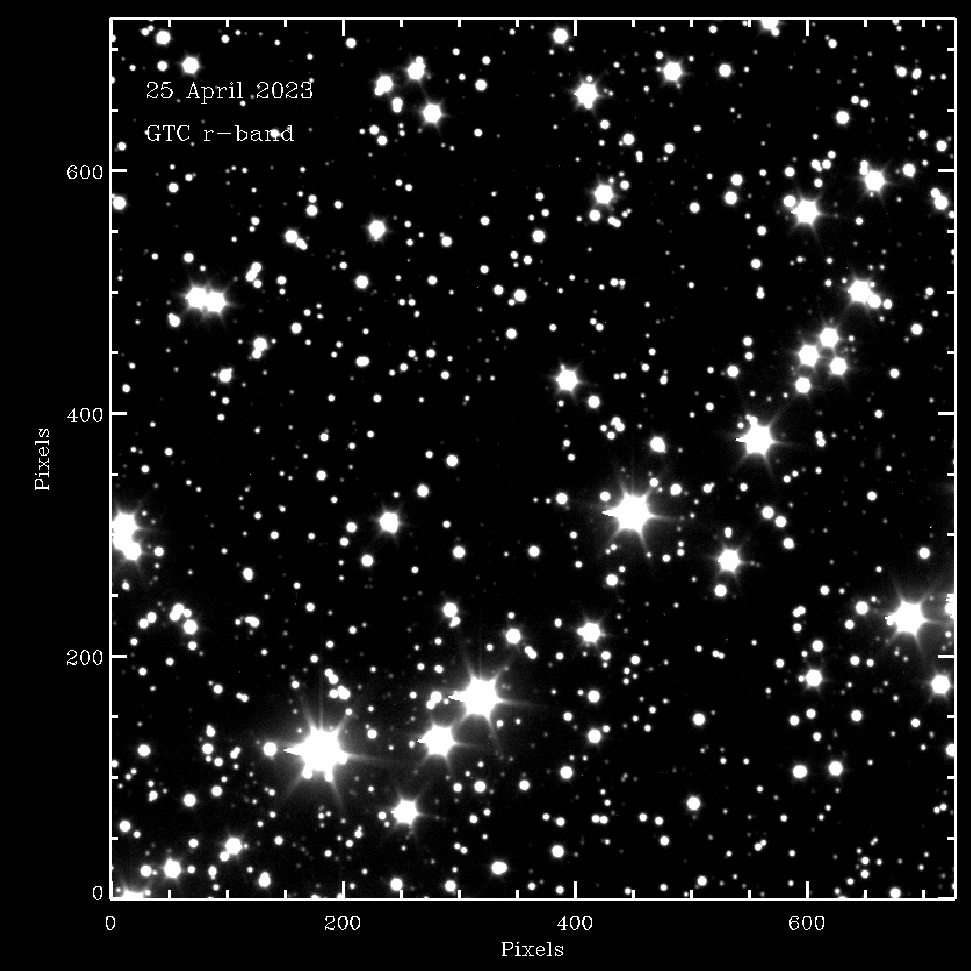}
\caption{Left: A 17\,\rm{min} exposure at $g$-band (left) and a 30\,\rm{min} exposure at $r$-band (right) with the 10.4-m GTC, centered on the triple transient. The 3x3\,\rm{arcmin} field is the same as the POSS I image in Fig.\,\ref{fig:SC}.  The triple transient is completely absent.  The limiting magnitude is $\sim$25.5 (3-sigma).}
\label{fig:GTC}
\end{figure*}

Fig.\,\ref{fig:GTC} shows the $r$ and $g$-band images obtained with the GTC, displayed with the same range of RA and DEC as the POSS I plate in Fig.\,\ref{fig:SC}, for direct comparison. Neither of the images obtained with the GTC show evidence of the triple transient.  


\begin{figure*}
\includegraphics[width=\textwidth]{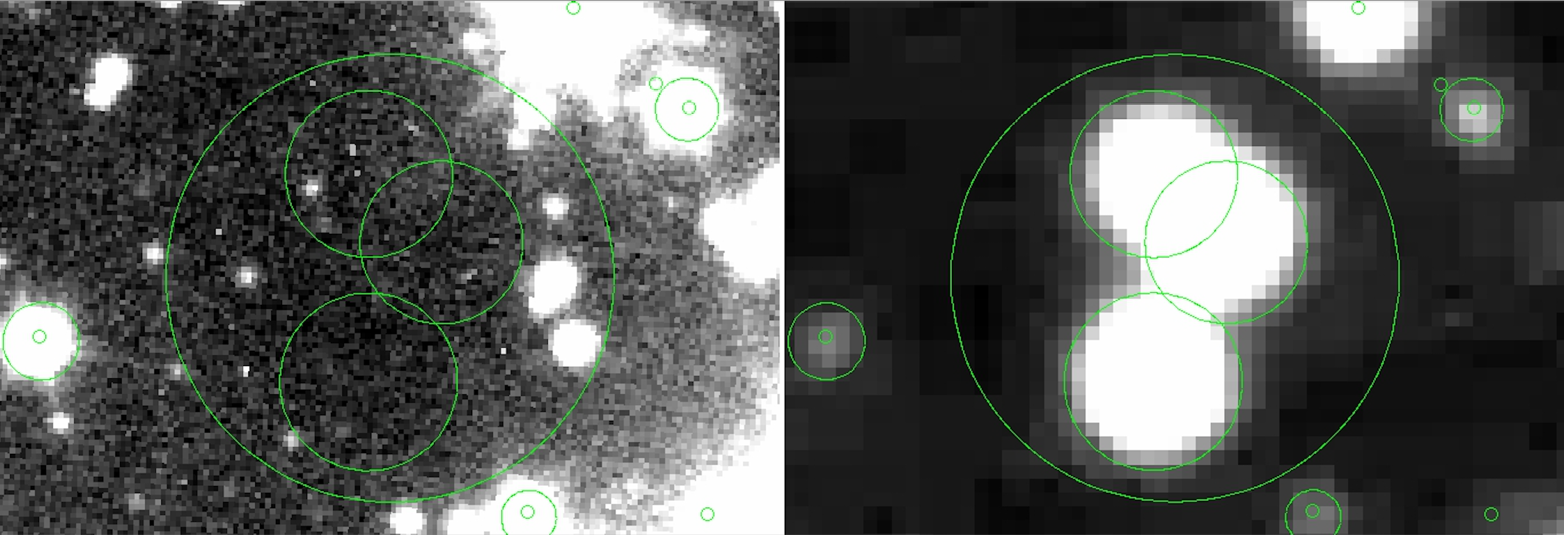}
\caption{An extreme zoom and gray-scale stretch of the 10.4-m GTC $r$-band image centered at the triple transient (left). The same zoom of the POSS I red image obtained in July 1952 is shown in the right panel. The positions of the three transients are marked by three large green circles, with astrometric calibration set by multiple neighbouring stars, some of which are marked by small circles. The GTC image exhibits no stars near the center of the three green circles, with a threshold of 25.5 mag. All the sources lying within the big green circle are below the POSS I detection limit and, thus, are not visible on the POSS-I (right) image.}
\label{fig:GTC_zoom}

\end{figure*}

Fig.\,\ref{fig:GTC_zoom} shows a zoom and extreme gray-scale stretch on the GTC $r$-band image (at left) that can be compared to the POSS I red image (right).  We place green circles around each of the transients in the POSS I image (at right), and we copy those three green circles onto the GTC image.  The green circles are positioned based on a careful astrometric calibration performed with numerous nearby stars to set their centers in RA and DEC.


Examining Fig.\,\ref{fig:GTC_zoom}, one sees that the the three 15th mag transients in POSS I do not appear in the green circles in the GTC r-band image.  But the GTC $r$-band image shows $\sim$25th mag stars located roughly 3 arcsec to the south and west of the centers of the northern and southern green circles.  These two 25th mag stars are not displaced by the same angular distance from the centers of their respective green circles, and there are many other point sources of ~25th mag in the image. Thus, it is entirely probable that these two 25th mag stars just happen, by chance, to be located ~3 arcsec from the previous location of the transients in the POSS-1 image.  At the expected location of the middle transient (green circle), there is no star in the GTC r-band image at all. This casts further doubt on stars displaced by 3 arcsec in the northern and southern green circles as being relevant to the original three transients from POSS-1. 

Also we counted the number of possible counterparts within three different radii (1, 3 respective 5 arcseconds) around each of the transients in POSS I and estimate the probability of finding a background object given the density of objects visible. This probability shows that the counterparts are unlikely to be physically connected to the transients.

\section{Analysis of the triple transient from the Palomar images}
\label{analysis}
We deblended and measured the brightness of the sources in the POSS I Red image from July 1952. We used \textsc{SExtractor} \citep{Bertin96} and \textsc{IRAF} \citep{Tody86}, separating the three objects as well as possible.  The resulting $R$ Supercosmos magnitudes of the sources measured by both \textsc{SExtractor} and \textsc{IRAF} as well as the approximate photometric uncertainty are given in Table \ref{tab_param}. We also estimated the $R$ magnitudes by measuring the FWHM of the profiles of six photometric reference stars within 10\,\rm{arcmin} and of the profiles of the three transient sources, yielding agreement with \textsc{SExtractor} and \textsc{IRAF} within 0.4\,mag, which is the value adopted as uncertainty. The uncertainty is due to the nonlinear photographic response to photons and to the overlap of the star-like sources. The total angular separation of the transients, from north to south, is 9.5\,\rm{arcsec}.

\begin{table}
\centering
\caption{Position and photometry of the triple transient}
\label{tab_param}
\begin{tabular}{llllll}
\hline 
Name & RA (J2000) & DEC (J2000) & R mag & R mag & Uncert. \\
 &  &  & (Sex.) & (IRAF) & \\
\hline 
Cand. 1 &
21:18:10.68 &
+50:22:46.85 &
16.0 &
15.7 &
0.4 \\
Cand. 2 &
21:18:10.35 &
+50:22:43.7 &
16.4 &
16.1 & 
0.4 \\
Cand. 3 & 
21:18:10.69 &
+50:22:37.45 &
14.9 &
15.1 & 
0.4 \\
\hline 
\end{tabular}
\end{table}

\begin{figure}
 \includegraphics[width=\columnwidth]{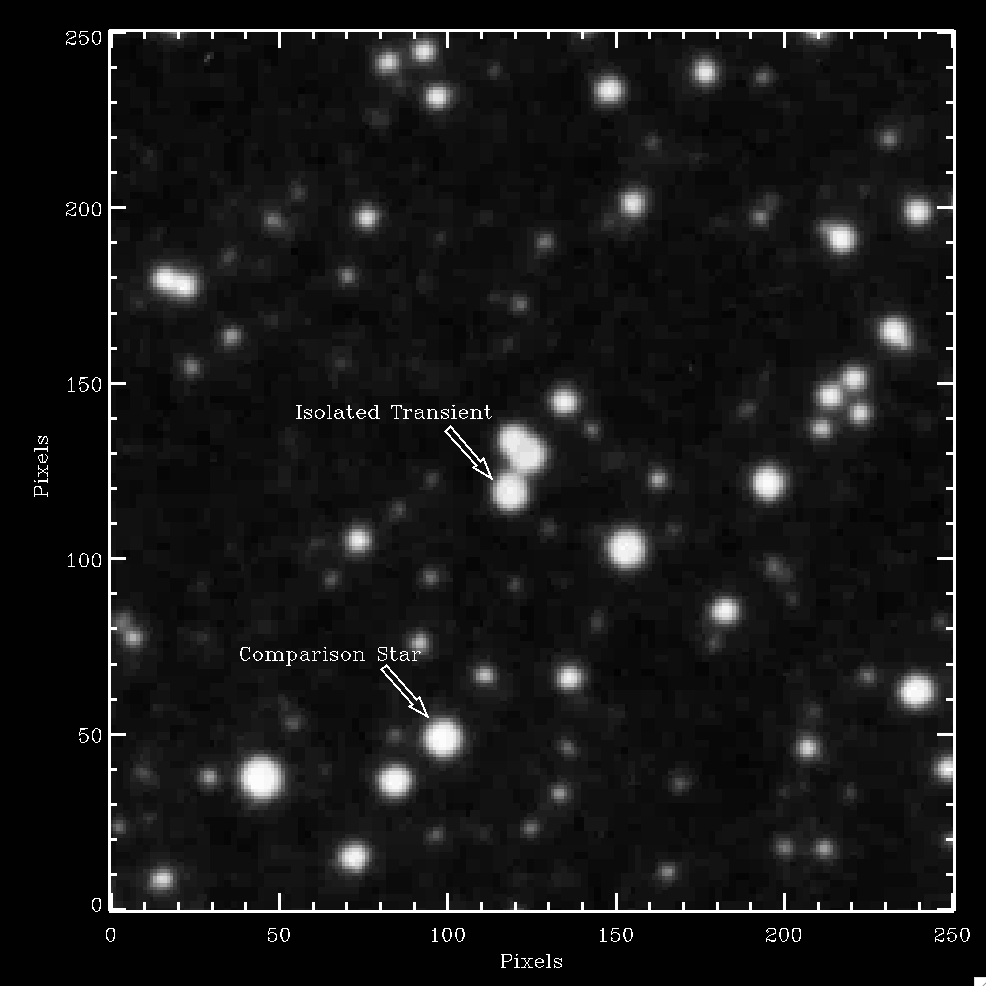}
    \caption{The POSS I red image, showing the transient that is nearly isolated from the other two. A comparison star is identified, having  similar brightness to serve as a proxy for the intrinsic profile of point sources resulting from the nonlinear response of the photographic plate.}
    \label{fig:compphot}
\end{figure}

Although a visual inspection already indicates that the transient has a clear circular shape with intensity diminishing radially outward, we compared the profile of the nearly isolated southern-most transient to that of a neighbouring star (Fig.\,\ref{fig:compphot}). We were especially attentive to the wings of the profile in both the north-south and east-west directions. If the transient exhibits qualitatively different profile shape at the flanks or wings of the profile, compared to the comparison star, its stellar-like nature may be questioned and its origin attributed to plate flaws or elementary particle hits.

\begin{figure}
 \includegraphics[width=\columnwidth]{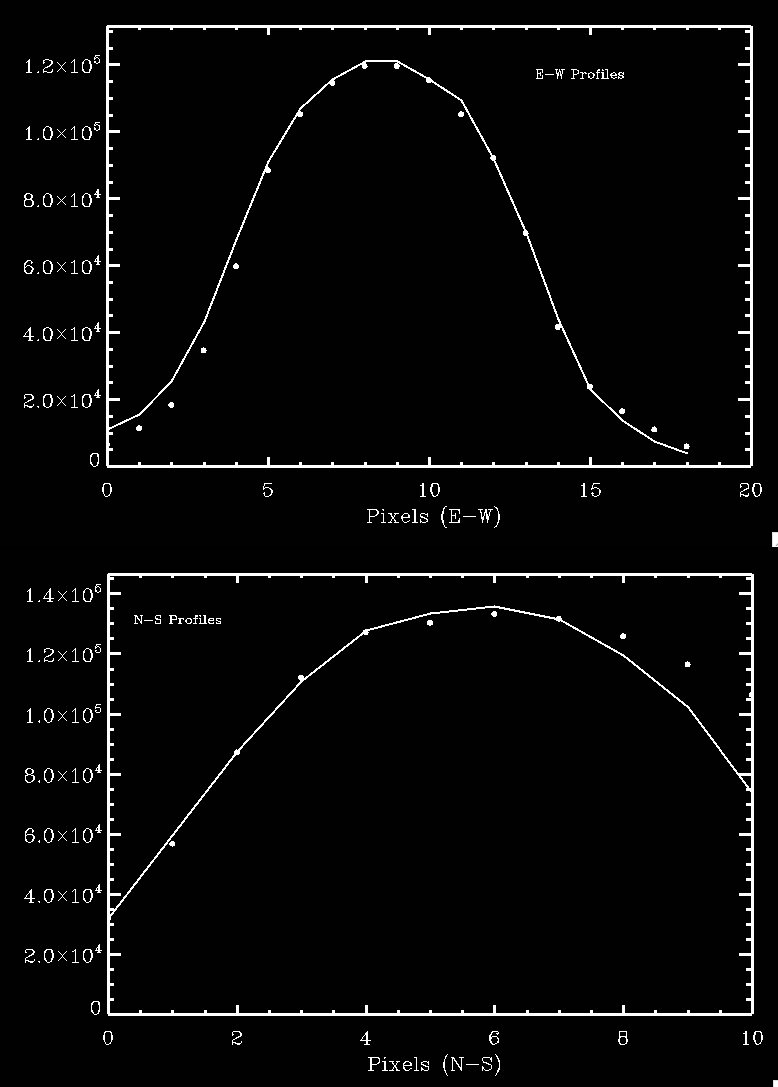}
    \caption{Comparison of the profiles of the isolated transient (dots) with a comparison star (solid line). Top: The east-west profiles (summed in N-S direction). Bottom: The north-south profiles (summed in E-W direction).}
    \label{fig:profile}
\end{figure}

Fig.\,\ref{fig:profile} shows the resulting profiles in the north-south and east-west directions. The profiles in both directions were produced by simply adding the digitized ADU.  The two profile shapes, of the transient and the comparison star, are remarkably similar, despite the magnitude of the two not being identical. The east-west profiles of the transient and comparison star reveal no discrepancy between the two. In the north-south direction the profiles are less similar due to the contamination of the transient located northward.  However, the level of agreement of the profiles in the north-nouth direction is good enough to discard a non-optical origin of the transient.  

In summary, we find no evidence that the transient is anything other than a bona fide unresolved, point source of light. In particular, the profiles show no evidence of a moving source such as an aircraft, asteroid, or elementary particle nor of a defect in the photographic plate.
Interestingly, the northern two transients have profiles that are slightly flat-topped compared to reference stars. This flat top is probably not due to saturation, as 16th mag stars are only slightly saturated in POSS I Red photographic images. This departure from the profiles of comparison stars is so slight that we can’t be sure if it is due to the blending of the two objects, or some sort of noise, or a real effect.

\section{Constraints on the size and distance}

The triple transient has equatorial coordinates, RA\,=\, 21$^{h}$18$^{m}$10.4$^{s}$, DEC\,=\,+50\textdegree22'43.4'' (J2000), placing it within 1 degree of the plane of the Milky Way, within the north-east region of Cygnus.  This location raises the possibility that the triple transient is somehow physically related to the plane of the Galaxy, which has a higher density of objects, including brown dwarfs, massive stars, white dwarfs, neutron stars, and black holes. This location of the triple transient within the galactic plane is consistent with the notion that a gravitational lens alignment is playing a role (Section~\ref{gl}).

\label{sizedist}
\subsection{Three independent sources}
We consider the possibility that the three transient sources are separate objects in space. Their physical properties are constrained by two observed properties, the dimming time within 50\,\rm{min} and their angular extent on the sky of 10 arcsec.
Space-time causality provides constraints on their size and distance, as follows.

The dimming time scale of 50\,\rm{min} is set by the two consecutive exposures in July 1952. The three point-sources delivered their photons with a fluence of $\sim$ 16th mag, each, during the 50 min exposure time.  The wavelength coverage of the red plate was 600 to 700 nm. But all three sources were absent in the 10\,\rm{min} POSS I blue exposure, between wavelengths 320 to 500 nm. This absence implies that they dimmed to roughly  
1\% of the brightness compared to the previous POSS I red exposure within its 50\,\rm{min}. The absence of the three sources in the two GTC exposures that reached magnitudes 25.5 imply that the sources are now apparently <1/10000 of peak brightness. The images from Pan-STARRS, SDSS, and ZTF confirm their absence.

The three sources could not have dimmed coincidentally and independently of each other as such dimming rarely occurs even in one source. Instead we may consider the possibility that the three sources brightened due to some common cause.  If so, the fastest that all three can be caused to brighten is the speed of light. If they were separated by more than 50 minutes of light-travel time, all three of them could not be "turned on" within the 50 minutes that they were observed to be "on". For the dimming of all three sources to be causally connected, the three sources must reside within a light-travel time, 50 minutes, of each other, corresponding to a distance less than s, given by:

\begin{equation}
    s < c \times 50min = 6\,au
\end{equation}
  
Thus, this mutual causality implies that the spacing between the three objects emitting the light must not differ by more than 6 au. However, the north and south point sources are separated by an angle on the sky of only $\sim$10\,\rm{arcsec}.  Assuming causal connection, the angular separation of 10 \rm{arcsec} and the maximum separation of 6 au implies a maximum distance to the triple source of 2 light-years.  Any greater distance, and the angular separation of 10 \rm arcsec would imply a physical separation more than 6 au, preventing a causal connection given the speed of light.
          
Therefore, to be causally connected, the three light sources must reside physically within 6 au of each other and are no more than 2 light-year away. This distance is less than the nearest star, the alpha Cen system, bringing the venue of the three transients to a distance within our Sun’s vicinity, if not the inner Solar System, or even Earth’s orbit. Such a close proximity, along with typical space velocities among stars in the solar neighborhood of $\sim$30 km s-1 \citep{Katz18} would imply accumulated proper motion of much more than 10 arcsec during 71 years, explaining why the three objects do not appear in the modern images.  Thus, the assumption of a shared cause of the dimming implies the three sources are close to our Solar System, including arbitrarily close. 

\subsection{Gravitational lensing}
\label{gl}

Alternatively, the three transients may be produced by a foreground gravitational lens having sufficient mass complexity to create image multiplicity, as commonly seen in strong gravitational lensing by intervening dark matter 
The rough sketch of a model consists of a single background source of optical light that is lensed by an intervening distribution of mass that produces, briefly, three nearly point-like images at Earth.  

Under this hypothesis, there are two possible scenarios. In the first scenario, a single background source is moving quickly enough perpendicular to the line-of-sight to be lensed by a multi-component mass distribution in the foreground, such as a planetary system around a star or a triple stellar system.  In this scenario, a sequence of chance alignments of that moving background source with different foreground mass concentrations causes a sequence of magnifications to produce the apparent three point-sources, one after the other, during the 50 min exposure.  One prediction of this scenario is that there must be a background star fainter than 25th magnitude, and also foreground masses of planets or stars that are also fainter than 25th magnitude, as no objects (foreground or background) brighter than 25th are observed in the two GTC images at r-band and g-band.  This scenario implies that all three gravitational lensing events in the sequence produce an optical brightening of at least 10,000x.  This scenario implies an unlikely sequence of nearly perfect alignments and enormous amplification for all three sources.  This seems improbable. 

In the second scenario, a background source intrinsically brightens at optical wavelengths by more than 10000x, presumably due to some explosive physical event. In the foreground there is a stationary, complex mass distribution that lenses that background explosion to produce three gravitational lens images that appear as three nearly point-like sources.  In this scenario, the optical emission from the explosive event is finished by the time the 50 min exposure is over, as all three point-sources fade to be undetectable in the POSS I blue image immediately after the POSS I red exposure of 50 min.  This scenario predicts that there may be a background source still there but now less that 0.01\% as bright as it was at peak luminosity, and there must be a foreground mass still there, also fainter than 25th mag. 

In this scenario, the three path lengths of light bent by the gravitational lens must not differ by more than $\sim$6 au, in order to keep all three rays of lensed light appearing within the 50 min of light-travel time. Foreground lens structures that cause path-length differences larger than ~6 au could not cause three point sources to naturally brighten and dim together within 50 min in a causal way.  This constraint rules out galactic-scale dark matter because the kiloparsec size of galaxies would yield path-length differences of many light years causing arrival-time differences of much longer than 50 minutes.

Novae, supernovae, and cataclysmic variables take many hours or days to brighten and dim, not within 50\,\rm{min}.  Flare stars do not become more luminous by factors of 10\,000.  One may consider the second scenario to take place inside or outside the Milky Way Galaxy.  If within the Galaxy, the peak luminosity would need to be comparable to the supergiant stars that are 14th mag from kiloparsecs away. For extragalactic sources, optical luminosities similar to those of supernovae, kilonovae, or gamma-ray bursts are required to explain the observed brightnesses here. To match the short time scale of the optical brightening, of less than 50\,\rm{min}, afterglows from mergers of compact objects may be considered.  However, the gravitational light-bending must not produce ray path differences more than 50\,\rm{min} in order to maintain coherent brightening within that time scale, ruling out galactic-scale mass distributions.   It is possible that the briefly luminous background source is outside the Milky Way, while the gravitational lens resides within the Milky Way. If so, this constitutes a method to detect condensed objects that are dark within the Milky Way or in its dark matter halo.

\section{Summary}

Three bright ($\sim$ 15th mag), optical point-like sources, appeared on the Palomar Observatory Sky Survey images taken on the 1952 of July 19. But within 1 hour they dimmed to a magnitude fainter than 21st magnitude at blue wavelengths, and they dimmed by more than 10 magnitudes, a factor of over 10,000, as seen in two non-detections with the 10.4-m GTC telescope.  The most isolated of the three sources permits measurement of its profile, which is indistinguishable from the profiles of stars in the same exposure.  


The observed time scale for the dimming of the triple transient of less than the exposure time of 50\,\rm{min} distinguishes these sources from common transients, notably supernovae and optical afterglow of gamma-ray bursts, that last for days and weeks,  \citep[e.g.,][]{Abdalla19, MAGIC19}. Also, SNe and GRBs are unresolved, single sources at extragalactic distances.  Here we clearly have three sources, not one.

We considered various scenarios to explain the factor of 10\,000 optical dimming of three point-sources with an angular separation on the sky of 10 arcsec. If they are separate objects, their dimming within 50\,\rm{min} constrains their physical separation in space to be less than 6 au for a causal connection. Their angular separation on the sky of only 10 arcsec, along with that 6 au maximum size, constrains the distance to the three objects to be less than 2 light year, and plausibly inside the Solar System.  Thus, one class of explanations involves three objects smaller than the Solar System and closer than the nearest star that brighten and dim within an hour.

Alternatively, the three sources are not separate physical objects but, instead, are images produced by a gravitational lens.  One may consider scenarios involving the gravitational lensing of a background explosive event that brightens and dims at optical wavelengths by 10,000x within 50\,\rm{min}. The foreground gravitational lens must have enough structure to produce three point-like images. We do not know what explosive events can provide the decrease in luminosity of 10,000x with a time scale of $\sim$ 1\,\rm{h}, as required by the observations.  Nor do we know what possible foreground mass distributions can produce three images. The sub-hour time-scale of the triple transient raises the question of a possible association with fast radio bursts (FRBs), events that can be associated to magnetars originated after the merger of two neutron stars \citep{Bochenek20, Moroianu23}. In this scenario, the FRB is in the background. It is likely that its optical luminosity increases by a process physically related to the process that causes the increase in luminosity in the radio regime. 

Similar dimming of multiple transient sources within a few arcmin, were reported by \citet{Villarroel21, Villarroel22arxiv}, along with similar dimming of thousands of transients by a factor of 10,000 \citep{Villarroel20, Solano22}. The nine transients in \citet{Villarroel21} within 10x10 arcmin raise the question of whether the same mechanism can explain the triple transient here, separated by only 10 arcsec. However, it is difficult to imagine a model that involves gravitational lensing that produces lensing images separated by 10 arcmin. 
The observation of the triple transient, as well as other similar events, calls for careful follow-up searches. 

\section*{Acknowledgements}
This research has made use of the Spanish Virtual Observatory (\url{https://svo.cab.inta-csic.es}) project funded by the Spanish Ministry of Science and Innovation/State Agency of Research MCIN/AEI/10.13039/501100011033 through grant PID2020-112949GB-I00 and MDM-2017-0737 at Centro de Astrobiología (CSIC-INTA). We also thank the teams and staff at the Gran Telescopio Canarias and at Space Laser Awareness for valuable technical and infrastructural help.  We thank the staff and technical teams at the Gran Telescopio Canarias for the excellent performance of the OSIRIS camera and 10.4-m GTC telescope facility. B.V. acknowledges Francisco Ricardo, Jacques Vallée, Eugene Jhong and Christopher Mellon for helpful discussions and support for VASCO research work by Swiss Philantropy Foundation and L’Oreal-UNESCO. She thanks Heterodox Academy for support via the Open Inquiry Courage Award.
The Pan-STARRS1 Surveys (PS1) and the PS1 public science archive have been made possible through contributions by the Institute for Astronomy, the University of Hawaii, the Pan-STARRS Project Office, the Max-Planck Society and its participating institutes, the Max Planck Institute for Astronomy, Heidelberg and the Max Planck Institute for Extraterrestrial Physics, Garching, The Johns Hopkins University, Durham University, the University of Edinburgh, the Queen’s University Belfast, the Harvard-Smithsonian Center for Astrophysics, the Las Cumbres Observatory Global Telescope Network Incorporated, the National Central University of Taiwan, the Space Telescope Science Institute, the National Aeronautics and Space Administration under Grant No. NNX08AR22G issued through the Planetary Science Division of the NASA Science Mission Directorate, the National Science Foundation Grant No. AST-1238877, the University of Maryland, Eotvos Lorand University (ELTE), the Los Alamos National Laboratory, and the Gordon and Betty Moore Foundation. This research has made use of data obtained from the SuperCOSMOS Science Archive, prepared and hosted by the Wide Field Astronomy Unit, Institute for Astronomy, University of Edinburgh, which is funded by the UK Science and Technology Facilities Council. This research has made use of the NASA/IPAC Infrared Science Archive, which is funded by the National Aeronautics and Space Administration and operated by the California Institute of Technology. This research has made use of IRAF. IRAF was distributed by the National Optical Astronomy Observatory, which was managed by the Association of Universities for Research in Astronomy (AURA) under a cooperative agreement with the National Science Foundation.This research has made use of Aladin (Bonnarel et al. 2000; Boch \& Fernique 2014) developed at CDS, Strasbourg Observatory, France. TOPCAT (Taylor 2005) and STILTS (Taylor 2006) have also been widely used in this paper.

\section*{Data Availability}\label{archive}

This paper is based mostly on publicly available data, including the Palomar Sky Survey from STScI and SuperCosmos, and publicly available data from Pan-STARRS, SDSS, ZTF, \textit{Gaia}, and USNO-B1.  In addition, we obtained two images with the 10.4-m GTC, $r$ and $g$-band, which are available from the first author upon request.




\bibliographystyle{mnras.bst} 
\bibliography{tripletransient} 




\bsp	
\label{lastpage}
\end{document}